\documentclass{article}
\usepackage[intlimits]{amsmath}
\usepackage[xdvi]{epsfig}
\usepackage[xdvi]{rotating}
\usepackage{epic,eepic}
\font\tenrm=cmr10 at 10pt

\DeclareMathOperator{\im}{Im}
\DeclareMathOperator{\re}{Re}
\DeclareMathOperator{\tr}{Tr}
\DeclareMathOperator{\arccot}{arccot}
\DeclareMathOperator{\artanh}{artanh}
\newcommand{\fslash}{\slash\!\!\!}
\newcommand{\ndot}{{\cdot}}
\numberwithin{equation}{section}

\begin{document}

\begin{titlepage}
\noindent
\begin{flushright}
\large
BONN-TH-97-05\\
hep-ph/9703368
\end{flushright}

\vspace*{3cm}

\begin{center}
{\huge
 \textbf{The Gerasimov-Drell-Hearn sum rule} \\[.2cm]
 \textbf{and the infinite-momentum limit}
}
\vspace{1cm}

{\Large
 Ralf Pantf\"order\footnote{e-mail: pantfoer@thew01.physik.uni-bonn.de},
 Horst Rollnik, and Walter Pfeil
}
\vspace{.5cm}

{\large
\textit {Physikalisches Institut der Universit\"at Bonn} \\
\textit {Nu{\ss}allee 12, 53115 Bonn, Germany}
}
\vspace{3cm}

\textbf {\large Abstract}

\end{center}

\begin{quote}
 We study the current-algebra approach to the Gerasimov-Drell-Hearn sum rule,
 paying particular attention to the infinite-momentum limit.
 Employing the ${\cal O}(\alpha^2)$ Weinberg-Salam model of weak interactions
 as a testing ground, we find that the legitimacy of the
 infinite-momentum limit is intimately
 connected with the validity of the naive equal-times algebra of electric
 charge densities. 
 Our results considerably reduce the reliability of a recently
 proposed modification of the Gerasimov-Drell-Hearn sum rule, 
 originating from an anomalous charge-density algebra.
\end{quote}

\end{titlepage}

\section{Introduction}

Exciting results on the spin structure of the proton derived from deep
inelastic scattering experiments with polarized muons and protons
brought the attention of particle physicists back to a rather old
issue: by general arguments of field theory and particle theory there
is a relation between a particle's anomalous magnetic moment and an
integral of a particular combination of the spin-dependent
photoabsorption cross sections of that particle.

For the nucleon, this relation is given by the Gerasimov-Drell-Hearn
(GDH) sum rule \cite{Gerasimov65,Drell66}, a direct experimental test
of which being in progress at several accelerators.  Yet
there are estimates taken in an indirect fashion from
pion-photoproduction data \cite
{Fox69,Karliner73,Workman92,Sandorfi94,Coersmeier93}.  Since the
integral in question runs over all photon energies, one has to
extrapolate the data, and even certain model assumptions about
multi-particle final states have to be made.  Nevertheless, a definite
discrepancy, particularly for the proton-neutron difference, remains
outside reasonable uncertainties, thus leading to a basic problem for
photo-hadron physics.

In view of this situation, various proposals to alter the GDH sum rule
have been published \cite {Kawarabayashi66c,Khare75,Chang94a}.  In
this article we point to a general difficulty which all of these
attempts encounter.

The GDH sum rule was originally derived from dispersion theory \cite
{Gerasimov65,Drell66}.  Conventional Regge phenomenology predicts the
dispersion integral to converge \cite{Mueller67b} (see also Ref.\
\cite{Ioffe84}), which has to do with \emph{moving} Regge poles only.
However, as Abarbanel and Goldberger \cite{Abarbanel68} pointed out, a
possible \emph{fixed} pole at angular momentum $J=1$ would modify the
sum rule by an additive constant -- essentially the residue of the
fixed pole.  We stress that it is by no means evident that the fixed
pole should be absent.  On the other hand, there is as yet no reliable
model prediction for the magnitude of its residue.

Alternatively, by the current-algebra approach, the GDH sum rule was
founded on the completeness sum in the infinite-momentum limit \cite
{Hosoda66a,Kawarabayashi66a}, assuming that the operators of electric
charge densities commute at equal times.  A few attempts to go beyond
this assumption can be found in the literature \cite
{Kawarabayashi66c,Khare75,Chang94a}.  However, these investigations
suffer from a severe deficiency which has not been noted before: the
infinite-momentum limit is handled in a naive way, the legitimacy of
which cannot be based on current algebra alone -- it enters as a mere
conjecture.

In the present article, we analyze the significance of the problems
connected with the infinite-momentum limit in some detail.  To this
end, we firstly derive the form that the GDH sum rule gets \emph
{without} taking the infinite-momentum limit. We call this equation
the finite-momentum GDH sum rule.  It is formulated in terms
of the (timelike) virtual forward Compton amplitude of the nucleon, or
any other fermion under consideration.  To examine the legitimacy of
the infinite-momentum limit, we need a perturbative model that allows
us to calculate the Compton amplitude for all values of the photon's
energy and virtuality.  Unfortunately, for the nucleon there is no
realistic model that works in the relevant kinematical domain.
Therefore, following Ref.\ \cite {Altarelli72}, we take the external
fermion to be an electron and employ the $\mathcal O(\alpha^2)$
Weinberg-Salam model, i.e.\ the standard model of electroweak
interactions of leptons to the fourth order of perturbation theory.
To this order, tree-level and one-loop Feynman graphs involving e$^-$,
$\nu_{\text e}$, $\gamma$, W$^\pm$, Z$^0$, and Higgs as internal
particles are to be worked out.

We obtain a non-vanishing charge-density commutator due to the
presence of the triangle anomaly, giving rise to a modification of the
finite-momentum GDH sum rule.  Our key result, however, is that if the
infinite-momentum limit is handled with care, it brings about a second
modification which exactly cancels the previous one, hence leaving the
(infinite-momentum) GDH sum rule unaltered.

We are thus led to the conclusion that the infinite-momentum limit has
to be regarded as being just as critical as the charge-density
commutator, especially in the presence of \emph {anomalous}
commutators. This fact greatly reduces the reliability of any proposed
modification of the GDH sum rule.

After some preliminaries in section 2, we review in section 3 the
dispersion-theoretic approach to the GDH sum rule.  Section 4 gives
the derivation of the finite-momentum GDH sum rule. Section 5
introduces the $\mathcal O(\alpha^2)$ Weinberg-Salam model and
presents the modifications brought about by charge-density algebra and
infinite-momentum limit.

\section{Preliminaries}

Throughout this paper, $M$, $Z$, and $\kappa$ denote the nucleon's mass,
charge (in units of $e$), and anomalous magnetic moment (in units of 
$\mu_{\text N}$), respectively, and $\alpha = e^2/4\pi$ is the 
fine-structure constant.
One-nucleon states $|p,\lambda\rangle$ with four-momentum $p$ and helicity
$\lambda=\pm1/2$ are normalized covariantly, $\langle p, \lambda
| p', \lambda' \rangle = (2\pi)^3\,2p^0\,\delta^3(\mbox{$\boldsymbol p -
\boldsymbol p'$})\,\delta_{\lambda\lambda'}$.  
Spinors $u(p,\lambda)$ are normalized as $\bar
u(p,\lambda)u(p,\lambda) = 2M\,\delta_{\lambda\lambda'}$.  Besides
this helicity basis, we use states $|p,s\rangle$ and spinors $u(p,s)$
with arbitrary spin four-vector $s$ obeying $s^2=-1$, $p\ndot s=0$,
and $\bar u(p,s)\gamma^\mu\gamma_5u(p,s) = 2Ms^\mu$.  Helicity
eigenstates are those for which the three-vectors $\boldsymbol p$ and
$\boldsymbol s$ are collinear.  For details we refer to the textbooks,
e.g.\ Ref.\ \cite [section 2-2] {Itzykson80}.  We adopt the
conventions $\epsilon^{123}=\epsilon^{0123}=+1$ and
$\gamma_5=i\gamma^0\gamma^1\gamma^2\gamma^3$.

We consider the virtual forward Compton amplitude
\begin{equation}
 T^{\mu\nu}(p,q,s) = i\!\int\!\text d^4x\, e^{iq\ndot x}
  \langle p, s| \text T J^\mu(x) J^\nu(0) | p, s \rangle.
\end{equation}
Its antisymmetric part has the invariant decomposition
\begin{equation} \label{inv-dec}
 \frac12\left(T^{\mu\nu} - T^{\nu\mu}\right)
  = -\frac iM\, \epsilon^{\mu\nu\rho\sigma} q_\rho s_\sigma
  A_1(\nu,q^2)
  -\frac i{M^3}\, \epsilon^{\mu\nu\rho\sigma} q_\rho
  \bigl( (M\nu) s_\sigma - (q\ndot s) p_\sigma \bigr) A_2(\nu,q^2),
\end{equation}
where $\nu = p\ndot q/M$ is the lab-frame energy of the photon.
In the current-algebra approach to the GDH sum rule, the linear combination
\begin{equation} \label{f2nuq2}
 f_2(\nu,q^2) = \frac\alpha{2M^2} \left(A_1(\nu,q^2) +
  \frac{q^2}{M\nu}A_2(\nu,q^2)\right)
\end{equation}
of the dimensionless invariant amplitudes $A_{1,2}(\nu,q^2)$ emerges
naturally.  For real photons we define as usual $f_2(\nu) \equiv f_2(\nu,0)$.

\section{Dispersion-theoretic approach to the GDH sum rule}

We now remind to the original dispersion-theoretic approach of refs.\
\cite {Gerasimov65,Drell66}, which assumes an \emph {unsubtracted}
dispersion relation for the forward Compton amplitude $f_2(\nu)$,
\begin{equation} \label{udr}
 \re{f_2(\nu)} = \frac2\pi\, \mathcal P\! \int_{\nu_0}^\infty\!
  \frac{\nu'\,\text d\nu'}{{\nu'}^2-\nu^2} \im{f_2(\nu')}.
\end{equation}
The constant $\nu_0 = m_\pi+m_\pi^2/2M$ is the pion-photoproduction
threshold, and $\mathcal P$ denotes principal value integration, which
can be omitted in case $\nu < \nu_0$.  Taking $\nu=0$ in Eq.\
\eqref{udr} yields
\begin{equation} \label{udr0}
 f_2(0) = \frac2\pi \!\int_{\nu_0}^\infty\! \frac{\text d\nu'}{\nu'}\,
  \im{f_2(\nu')}.
\end{equation}
The GDH sum rule
\begin{equation} \label{gdh}
 -\frac{2\pi^2\alpha\kappa^2}{M^2} =
  \int_{\nu_0}^\infty\! \frac{\text d\nu}\nu
  \bigl(\sigma_{1/2}(\nu) - \sigma_{3/2}(\nu)\bigr)
\end{equation}
is obtained from Eq.\ \eqref{udr0} by using the low-energy theorem of Low
\cite{Low54} and Gell-Mann and Goldberger \cite{Gell-Mann54},
\begin{equation}
 f_2(0) = -\frac{\alpha\kappa^2}{2M^2},
\end{equation}
and the optical theorem for the imaginary part of the forward amplitude,
\begin{equation} \label{op-th}
 8\pi \im f_2(\nu) = \sigma_{1/2}(\nu) - \sigma_{3/2}(\nu).
\end{equation}
Here, $\sigma_{1/2}(\nu)$ and $\sigma_{3/2}(\nu)$ denote the photoabsorption
cross sections of the nucleon for total photon-nucleon helicities 1/2 and 3/2,
respectively.

The validity of the unsubtracted dispersion relation \eqref{udr}
requires not only the \emph{imaginary} part of $f_2(\nu)$ to vanish
sufficiently rapid at large $\nu$ in order that the integral will
converge. Besides, the \emph{real} part has to vanish too, which means
$f_2(\infty)=0$.  That this need not necessarily be the case was shown
by Abarbanel and Goldberger \cite{Abarbanel68}.  In Regge language, a
possible non-vanishing $f_2(\infty)$ is equivalent to a $J=1$ fixed
pole in angular-momentum plane.

The easiest way to see what modification is brought about by
a non-vanishing $f_2(\infty)$,
is to write down a \emph{subtracted} dispersion relation,
\begin{equation} \label{sdr}
 \re{f_2(\nu)} - f_2(0) = \frac2\pi\, \mathcal P \!\int_{\nu_0}^\infty\!
  \frac{\nu^2\,\text d\nu'}{\nu'({\nu'}^2-\nu^2)} \im{f_2(\nu')}.
\end{equation}
Letting $\nu$ approach \emph{infinity} now, one gets
\begin{equation} \label{sdr0}
 f_2(0) = \frac2\pi \!\int_{\nu_0}^\infty\! \frac{\text d\nu'}{\nu'}\,
  \im{f_2(\nu')} + f_2(\infty).
\end{equation}
This gives rise to a finite modification of the GDH sum rule.  Note that the
subtraction here was \emph{not} enforced by a divergent integral, in which
case it would have been impossible to drag the limit $\nu\to\infty$ inside
the $\nu'$ integral in Eq.\ \eqref{sdr}.

We emphasize that to our knowledge, there is \emph{no
fundamental reason} requesting the constant $f_2(\infty)$ to vanish!

\section{Current-algebra approach -- the finite-mo\-men\-tum GDH sum rule}

In this section we remind the reader to the current-algebra derivation
of the GDH sum rule, which is based essentially on two premises.  Firstly,
electric charge densities are assumed to commute at equal times.
Secondly, one assumes that taking the infinite-momentum limit is
legitimate.  We stress that there are ans\"atze \cite
{Kawarabayashi66c,Khare75,Chang94a} that weaken the former
assumption, but the latter one has never been questioned seriously.

We follow the idea of Hosoda and Yamamoto \cite{Hosoda66a}, but
we postpone the infinite-momentum limit to the very end of the calculation
in order to shed some light on its meaning.
For the sake of transparency, we explicitely write down some formulae
which are known from the literature, e.g.\ Ref.\ \cite{Itzykson80}.

From causality arguments, the equal-times commutator
$[J^0(\boldsymbol x, 0), J^0(\boldsymbol y, 0)]$ of electric charge
densities must be a finite sum over derivatives of the delta function
$\delta^3(\boldsymbol x - \boldsymbol y)$.
Here we start from the \emph{naive} commutator
\begin{equation} \label{naive-comm}
 [J^0(\boldsymbol x, 0), J^0(\boldsymbol y, 0)] = 0,
\end{equation}
which can formally be obtained by writing $J^0(x) = \sum_{\text f}
Z_{\text f}^{} q_{\text f}^\dagger(x)q_{\text f}^{}(x)$ and
employing canonical anticommutation relations among
quark fields $q_{\text f}^\dagger(x)$, $q_{\text f}(x)$.  We define
the operator of the electric dipole moment as usual,
\begin{equation}
 D^i(x^0) = e \!\int\! \text d^3x\,x^i J^0(x),
\end{equation}
and sandwich the commutator of components $D^\pm(0) \equiv
(D^1(0) \pm iD^2(0))/\sqrt2$ between one-nucleon states of positive 
helicity, taking the incoming nucleon to be traveling along the
$x^3$-axis, $p^\mu = (p^0, 0, 0, \sqrt{(p^0)^2 - M^2})$,
\begin{equation} \label{naive-comm-me}
 \langle p', \tfrac12 | [D^+(0), D^-(0)] | p, \tfrac12 \rangle = 0.
\end{equation}

We now insert a complete set of intermediate states and separate the
one-nucleon states from the continuum.

In terms of form factors $F_{1,2}(q^2)$ with normalization $F_1(0)=Z$,
$F_2(0)=\kappa$, the one-nucleon matrix element of the dipole-moment
operator reads
\begin{equation}
 \langle k, \lambda | D^i(0) | p, \tfrac12 \rangle =
  ie\,(2\pi)^3\, \nabla^i \delta^3(\boldsymbol q)\, u^\dagger(k, \lambda)
  \left( F_1(q^2) + \boldsymbol\gamma\ndot\boldsymbol q\, \frac{F_2(q^2)}{2M}
  \right) u(p,\tfrac12),
\end{equation}
where $q=k-p$.  Therewith, the one-nucleon 
intermediate-state contribution to the matrix element \eqref{naive-comm-me}
is obtained straightforwardly,
\begin{align} \label{comm-1n}
 \langle p', \tfrac12 | [D^+(0), D^-(0)] | p, 
       \tfrac12 \rangle_{\text{one-nucleon}}
 & = \sum_{\lambda=\pm\frac12}
     \!\int\!\frac{\text d^3 k}{(2\pi)^3 2k^0}\, 
     \langle p', \tfrac12 | D^+(0) | k, \lambda \rangle
     \langle k, \lambda | D^-(0) |  p, \tfrac12 \rangle \notag\\
 & \quad {} - \{+\leftrightarrow-\}\notag\\
 & = (2\pi)^3\, 2p^0\, \delta^3(\boldsymbol p' - \boldsymbol p)
     \left( \frac{2\pi\alpha\kappa^2}{M^2} -
     \frac{2\pi\alpha(Z+\kappa)^2}{(p^0)^2}\right).
\end{align}
Here we stress the presence of the second term, which vanishes in the 
infinite-momentum limit.  Hitherto, its only appearance in the literature 
was in Ref.\ \cite{Pradhan72}, where it was incorrect.  

On the other hand, the
continuum contribution, i.e.\ the sum over all intermediate states
$|\text I\rangle$ \emph{except} the one-nucleon state, can be obtained 
by virtue of current conservation $\partial_\mu J^\mu(x) = 0$, which
implies $\dot D^i(x^0) = e \!\int\! \text d^3x J^i(x)$, and using
translational invariance to carry out the spatial integrations,
\begin{align}
 \langle p', \tfrac12 | [D^+(0), D^-(0)] | p, \tfrac12 \rangle_{\text{cont}}
  & = {\sum_{\text I}}'
  \langle p', \tfrac12 | D^+(0) | \text I \rangle
  \langle \text I | D^-(0) |  p, \tfrac12 \rangle
  - \{+\leftrightarrow-\}\notag\\
 & = (2\pi)^3\, \delta^3(\boldsymbol p' - \boldsymbol p)\,
  {\sum_{\text I}}'
  (2\pi)^3\, \delta^3(\boldsymbol p_{\text I} - \boldsymbol p)\,
  \frac{4\pi\alpha
  |\langle p, {\tfrac12} | J^+(0) | \text I \rangle |^2}
  {(p^0-p_{\text I}^0)^2}\notag\\
 & \quad {}- \{+\to-\}.
\end{align}
The $\pm$ components of the current are defined by
$J^\pm(x)\equiv(J^1(x)\pm i J^2(x))/\sqrt2$.  Introducing the timelike
virtual photon momentum $q$ with $\boldsymbol q = 0$, we can
substitute $
 \delta^3(\boldsymbol p_{\text I} - \boldsymbol p) =
  \int_{q^0_{\fam0{thr}}}^\infty \!\text d q^0\,\delta^4(p_{\text I} - p - q)$,
where the pion-production threshold
$q^0_{\text{thr}} \equiv M\nu_{\text{thr}}/p^0$
is determined by $(p+q_{\text{thr}})^2 = (M+m_\pi)^2$.
For $p^0\to\infty$, $\nu_{\text{thr}}$ approaches the familiar
pion-photoproduction threshold $\nu_0 = m_\pi + m_\pi^2/2M$.

We can now express the continuum contribution in terms of the forward
virtual Compton amplitude $f_2(\nu,q^2)$, Eq.\ \eqref{f2nuq2},
\begin{align} \label{comm-cont}
 \langle p', \tfrac12 | [D^+(0), D^-(0)] | p, \tfrac12 \rangle_{\text{cont}}
 & = (2\pi)^3\, 2p^0\, \delta^3(\boldsymbol p' - \boldsymbol p)
  \!\int_{q^0_{\text{thr}}}^\infty \!\frac{\text d q^0}{q^0}\,
  \frac\alpha{p^0q^0} \!\int\!\text d^4 x\, e^{iqx}
  \langle p, {\tfrac12}|J^+(x)J^-(0)|p,{\tfrac12}\rangle
  \notag\\
 & \quad {}- \{+\leftrightarrow-\}\notag\\
 & = (2\pi)^3\, 2p^0\, \delta^3(\boldsymbol p' - \boldsymbol p) \,
  8\!\!\int_{\nu_{\text{thr}}}^\infty\! \frac{\text d \nu}{\nu}\,
  \im f_2\!\left(\nu, \frac{M^2\nu^2}{(p_0)^2}\right).
\end{align}

\begin{figure}[tb]
 \begin{center}
  \input{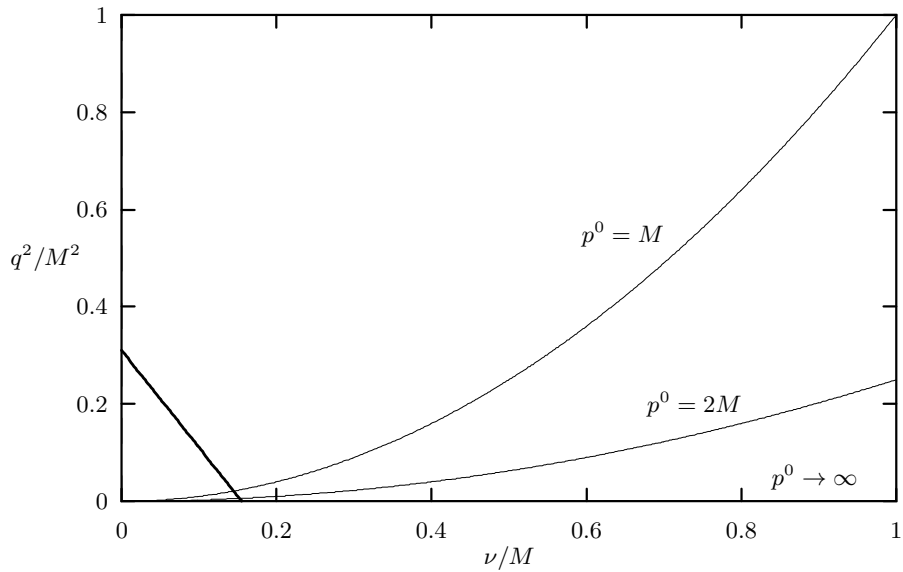}
 \end{center}
 \caption{\footnotesize
  The integration path of the finite-momentum GDH sum rule
  \eqref{fmgdh} in the $(\nu,q^2)$ plane for nucleon energy $p^0=M$ and
  $p^0=2M$.  The heavy line represents the pion-production threshold.
  \label{q2nu}}
\end{figure}

Since the one-nucleon part \eqref{comm-1n} and the continuum
part \eqref{comm-cont} sum up to give the commutator matrix element
\eqref{naive-comm-me}, we conclude
\begin{equation} \label{fmgdh}
 -\frac{2\pi^2\alpha\kappa^2}{M^2} + \frac{2\pi^2\alpha(Z+\kappa)^2}{(p^0)^2} =
  \int_{\nu_{\text{thr}}}^\infty\! \frac{\text d \nu}\nu\,
  8\pi \im f_2\!\left(\nu, \frac{M^2\nu^2}{(p^0)^2}\right).
\end{equation}
We call this equation the \emph{finite-momentum} GDH sum
rule. It is based solely on the naive charge-density commutator
\eqref{naive-comm}, or on the weaker assumption presented by Eq.\
\eqref{naive-comm-me}.  In particular, the integral on the right-hand side
of Eq.\ \eqref{fmgdh} \emph{converges}, since it relies only on the
validity of the completeness relation for the physical intermediate
states.  This is irrespective of the convergence of the genuine GDH
integral with its integrand $8\pi\im f_2(\nu,0)/\nu$.  The integration
path in the $(\nu,q^2)$ plane for various values of the energy $p^0$
is depicted in Fig.\ \ref{q2nu}.  Note that for any finite value of
$p^0$ this path is a parabola that extends to arbitrarily high
timelike photon virtualities.%

Taking the infinite-momentum limit now constitutes the last but one
step of the derivation of the GDH sum rule,
\begin{equation} \label{pre-gdh}
 -\frac{2\pi^2\alpha\kappa^2}{M^2} =
  \lim_{p^0\to\infty} \int_{\nu_{\text{thr}}}^\infty\!
  \frac{\text d \nu}\nu\, 8\pi \im f_2\!\left(\nu, \frac{M^2\nu^2}{(p^0)^2}
  \right).
\end{equation}
To get the usual form of the sum rule one now has to interchange the
limit $p^0\to\infty$ with the $\nu$ integration.  If the properties of
the function $\im f_2(\nu,q^2)$ allow the limit to be dragged into the
integral, the GDH sum rule follows from Eq.\ \eqref{pre-gdh} with the
help of the optical theorem \eqref{op-th}.  In principle, however,
permuting limit and integration could give rise to a (finite or
infinite) modification of the sum rule.  We stress that current
algebra has no answer to this problem.

\begin{figure}[tb]
 \center
  \psfig{file=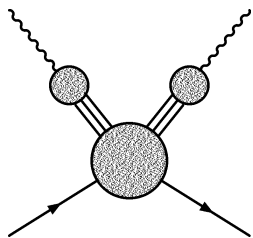,bbllx=5cm,bburx=14cm,bblly=25cm,bbury=29cm}
 \caption[]{\footnotesize
  Intermediate hadron states in virtual Compton scattering,
  leading to singularities in the photon mass $q^2$.
  \label{int-hadron}
 }
\end{figure}

Nevertheless, one can easily see the origin of possible difficulties
on quite general grounds.
We recall that the timelike virtual Compton amplitude meets singularities
in the photon mass $q^2$ due to intermediate hadron states, as indicated
in Fig.\ \ref{int-hadron}.
Thus one expects
for the amplitude $f_2(\nu,q^2)$ a spectral representation of the form
\begin{equation}  \label{spec-rep}
 \im f_2(\nu,q^2) = \frac1\pi \int_{q_0^2}^\infty\!
  \frac{\text dq^{\prime2}}{q^2 - q^{\prime2}}\, \rho(\nu,q^{\prime2}),
\end{equation}
where $q_0^2$ is the mass of the lowest-lying state that couples to the
photon.  Inserting Eq.\ \eqref{spec-rep} into Eq.\ \eqref{pre-gdh}, it
is evident that for finite $p^0$, the $\nu$ integration also meets the
$q^2$ singularities, since $q^2=M^2\nu^2/(p^0)^2$.
Only if one can drag the limit $p^0\to\infty$ inside the integral in Eq.\
\eqref{pre-gdh}, these singularities play no explicite role.

In the following section we will show, by adopting a concrete model,
that indeed non-trivial modifications of the current-algebra arguments
are to be expected.

\section{The finite-momentum GDH sum rule within the
 $\boldsymbol{\mathcal O(\alpha^2)}$ Weinberg-Salam model}

We now examine the finite-momentum GDH sum rule and the
infinite-mo\-men\-tum limit for the Compton amplitude of the electron
within the $\mathcal O(\alpha^2)$ Weinberg-Salam model.  Let $m$ and
$M_{\text Z}$ denote the mass of the electron and the Z$^0$ boson,
respectively.  For simplicity, we will take the Weinberg angle
$\theta_{\text W}$ to be such that the coupling of the Z$^0$ boson to
electrons is purely axial-vector, i.e.\ $\sin^2{\theta_{\text W}} =
1/4$.  The Fermi constant is then given by $G_{\text F}/\sqrt2 =
8\pi\alpha/3M_{\text Z}^2$.  We will expand in the coupling constant
$e$ only, regarding all masses as given parameters, thus $m^2G_{\text
F}$ will be of order $\alpha$.

\begin{figure}[tb]
 \center
  \psfig{file=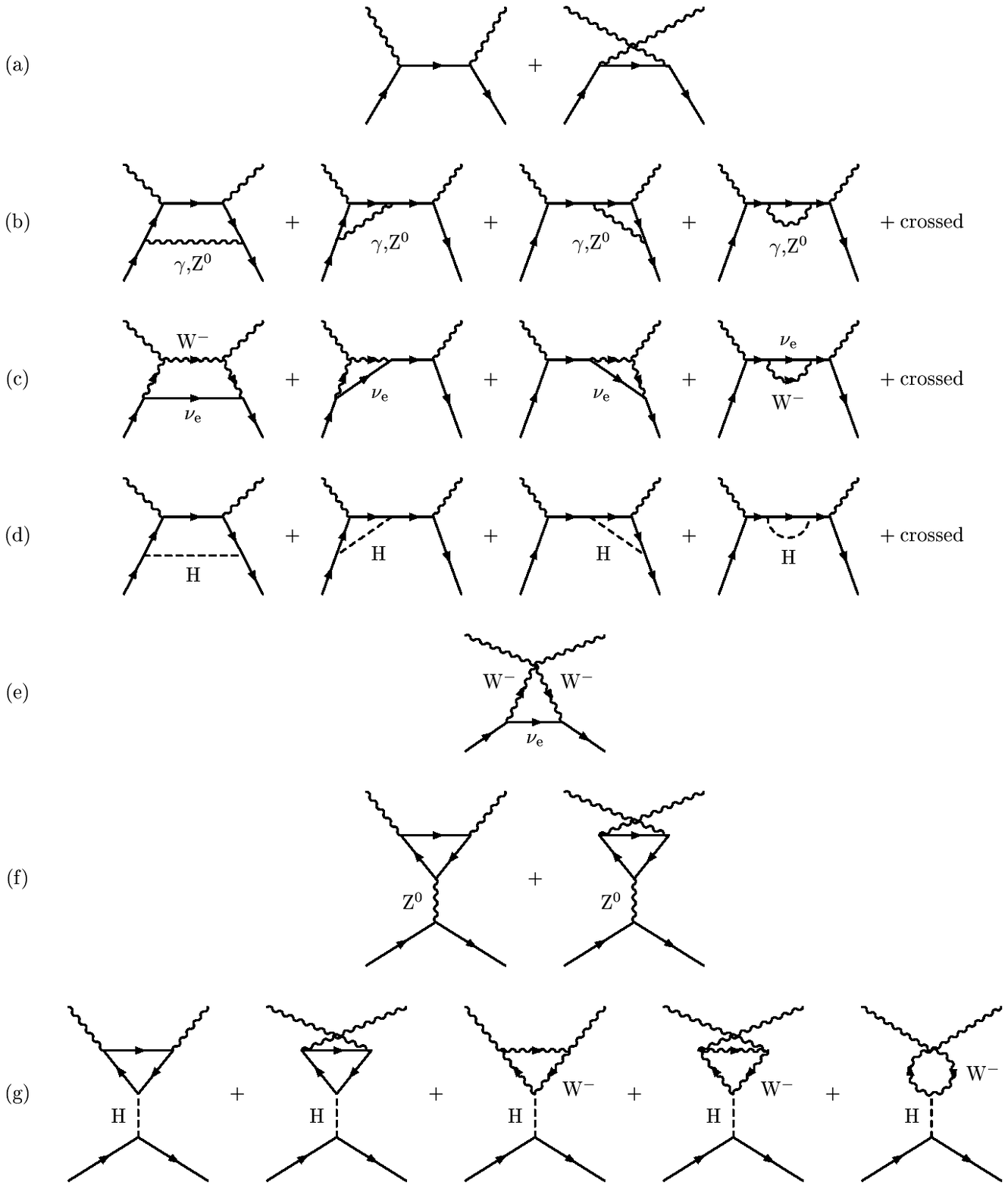,
	 bbllx=0cm,bburx=15cm,bblly=6cm,bbury=23cm}
 \caption[]{\footnotesize
  Feynman graphs contributing to the forward Compton
  amplitude of the $\mathcal O(\alpha^2)$ Weinberg-Salam model;
  solid lines represent e$^-$ and $\nu_{\text e}$,
  wavy lines are the gauge bosons $\gamma$, Z$^0$, and W$^\pm$,
  and a dashed line is the Higgs boson. \\
  \hspace*{1em} (a) tree graphs \\
  \hspace*{1em} (b) e$^-\gamma$ and e$^-$Z$^0$ intermediate states \\
  \hspace*{1em} (c) $\nu_{\text e}$W$^-$ intermediate state \\
  \hspace*{1em} (d) e$^-$H intermediate state \\
  \hspace*{1em} (e) WW$\gamma\gamma$ contact graph \\
  \hspace*{1em} (f) Higgs exchange \\
  \hspace*{1em} (g) Z$^0$ exchange \\
  The graphs with crossed external photon lines are omitted in (b)--(d).
  Not shown are external line insertions such as vacuum polarization.
  \label{graphs}
 }
\end{figure}

\begin{figure}[tb]
 \begin{center}
\setlength{\unitlength}{0.240900pt}
\begin{picture}(1500,900)(0,0)
\tenrm
\thinlines \drawline[-50](220,495)(1436,495)
\thicklines \path(220,113)(240,113)
\thicklines \path(1436,113)(1416,113)
\put(198,113){\makebox(0,0)[r]{\small $-60$}}
\thicklines \path(220,240)(240,240)
\thicklines \path(1436,240)(1416,240)
\put(198,240){\makebox(0,0)[r]{\small $-40$}}
\thicklines \path(220,368)(240,368)
\thicklines \path(1436,368)(1416,368)
\put(198,368){\makebox(0,0)[r]{\small $-20$}}
\thicklines \path(220,495)(240,495)
\thicklines \path(1436,495)(1416,495)
\put(198,495){\makebox(0,0)[r]{\small $0$}}
\thicklines \path(220,622)(240,622)
\thicklines \path(1436,622)(1416,622)
\put(198,622){\makebox(0,0)[r]{\small $20$}}
\thicklines \path(220,750)(240,750)
\thicklines \path(1436,750)(1416,750)
\put(198,750){\makebox(0,0)[r]{\small $40$}}
\thicklines \path(220,877)(240,877)
\thicklines \path(1436,877)(1416,877)
\put(198,877){\makebox(0,0)[r]{\small $60$}}
\thicklines \path(220,113)(220,133)
\thicklines \path(220,877)(220,857)
\put(220,68){\makebox(0,0){\small $0.001$}}
\thicklines \path(423,113)(423,133)
\thicklines \path(423,877)(423,857)
\put(423,68){\makebox(0,0){\small $0.01$}}
\thicklines \path(625,113)(625,133)
\thicklines \path(625,877)(625,857)
\put(625,68){\makebox(0,0){\small $0.1$}}
\thicklines \path(828,113)(828,133)
\thicklines \path(828,877)(828,857)
\put(828,68){\makebox(0,0){\small $1$}}
\thicklines \path(1031,113)(1031,133)
\thicklines \path(1031,877)(1031,857)
\put(1031,68){\makebox(0,0){\small $10$}}
\thicklines \path(1233,113)(1233,133)
\thicklines \path(1233,877)(1233,857)
\put(1233,68){\makebox(0,0){\small $100$}}
\thicklines \path(1436,113)(1436,133)
\thicklines \path(1436,877)(1436,857)
\put(1436,68){\makebox(0,0){\small $1000$}}
\thicklines \path(220,113)(1436,113)(1436,877)(220,877)(220,113)
\put(45,495){\makebox(0,0)[l]{\shortstack{\small \begin{sideways} 		$\big(\sigma_{1/2}(\nu) - \sigma_{3/2}(\nu)\big)/{\rm mbarn}$ 		\end{sideways}}}}
\put(828,23){\makebox(0,0){\small $\nu/m$}}
\thinlines \path(220,491)(220,491)(232,490)(245,489)(257,489)(269,488)(281,487)(294,485)(306,484)(318,482)(331,480)(343,478)(355,476)(367,473)(380,470)(392,466)(404,462)(417,457)(429,452)(441,446)(453,439)(466,431)(478,423)(490,413)(503,402)(515,390)(527,377)(539,362)(552,346)(564,329)(576,311)(588,291)(601,271)(613,251)(625,230)(638,209)(650,190)(662,172)(674,157)(687,145)(699,138)(711,136)(724,140)(736,150)(748,167)(760,190)(773,220)(785,257)(797,298)(810,343)(822,391)
\thinlines \path(822,391)(834,440)(846,490)(859,538)(871,584)(883,626)(896,664)(908,698)(920,726)(932,750)(945,768)(957,781)(969,790)(982,794)(994,795)(1006,792)(1018,787)(1031,779)(1043,770)(1055,759)(1068,747)(1080,734)(1092,721)(1104,707)(1117,694)(1129,680)(1141,667)(1153,654)(1166,642)(1178,631)(1190,620)(1203,609)(1215,599)(1227,590)(1239,582)(1252,574)(1264,567)(1276,560)(1289,554)(1301,548)(1313,543)(1325,538)(1338,534)(1350,530)(1362,526)(1375,523)(1387,520)(1399,518)(1411,515)(1424,513)(1436,511)
\thicklines \path(220,495)(220,491)
\thicklines \path(232,495)(232,490)
\thicklines \path(245,495)(245,489)
\thicklines \path(257,495)(257,489)
\thicklines \path(269,495)(269,488)
\thicklines \path(281,495)(281,487)
\thicklines \path(294,495)(294,485)
\thicklines \path(306,495)(306,484)
\thicklines \path(318,495)(318,482)
\thicklines \path(331,495)(331,480)
\thicklines \path(343,495)(343,478)
\thicklines \path(355,495)(355,476)
\thicklines \path(367,495)(367,473)
\thicklines \path(380,495)(380,470)
\thicklines \path(392,495)(392,466)
\thicklines \path(404,495)(404,462)
\thicklines \path(417,495)(417,457)
\thicklines \path(429,495)(429,452)
\thicklines \path(441,495)(441,446)
\thicklines \path(453,495)(453,439)
\thicklines \path(466,495)(466,431)
\thicklines \path(478,495)(478,423)
\thicklines \path(490,495)(490,413)
\thicklines \path(503,495)(503,402)
\thicklines \path(515,495)(515,390)
\thicklines \path(527,495)(527,377)
\thicklines \path(539,495)(539,362)
\thicklines \path(552,495)(552,346)
\thicklines \path(564,495)(564,329)
\thicklines \path(576,495)(576,311)
\thicklines \path(588,495)(588,291)
\thicklines \path(601,495)(601,271)
\thicklines \path(613,495)(613,251)
\thicklines \path(625,495)(625,230)
\thicklines \path(638,495)(638,209)
\thicklines \path(650,495)(650,190)
\thicklines \path(662,495)(662,172)
\thicklines \path(674,495)(674,157)
\thicklines \path(687,495)(687,145)
\thicklines \path(699,495)(699,138)
\thicklines \path(711,495)(711,136)
\thicklines \path(724,495)(724,140)
\thicklines \path(736,495)(736,150)
\thicklines \path(748,495)(748,167)
\thicklines \path(760,495)(760,190)
\thicklines \path(773,495)(773,220)
\thicklines \path(785,495)(785,257)
\thicklines \path(797,495)(797,298)
\thicklines \path(810,495)(810,343)
\thicklines \path(822,495)(822,391)
\thicklines \path(834,495)(834,440)
\thicklines \path(846,495)(846,490)
\thicklines \path(859,495)(859,538)
\thicklines \path(871,495)(871,584)
\thicklines \path(883,495)(883,626)
\thicklines \path(896,495)(896,664)
\thicklines \path(908,495)(908,698)
\thicklines \path(920,495)(920,726)
\thicklines \path(932,495)(932,750)
\thicklines \path(945,495)(945,768)
\thicklines \path(957,495)(957,781)
\thicklines \path(969,495)(969,790)
\thicklines \path(982,495)(982,794)
\thicklines \path(994,495)(994,795)
\thicklines \path(1006,495)(1006,792)
\thicklines \path(1018,495)(1018,787)
\thicklines \path(1031,495)(1031,779)
\thicklines \path(1043,495)(1043,770)
\thicklines \path(1055,495)(1055,759)
\thicklines \path(1068,495)(1068,747)
\thicklines \path(1080,495)(1080,734)
\thicklines \path(1092,495)(1092,721)
\thicklines \path(1104,495)(1104,707)
\thicklines \path(1117,495)(1117,694)
\thicklines \path(1129,495)(1129,680)
\thicklines \path(1141,495)(1141,667)
\thicklines \path(1153,495)(1153,654)
\thicklines \path(1166,495)(1166,642)
\thicklines \path(1178,495)(1178,631)
\thicklines \path(1190,495)(1190,620)
\thicklines \path(1203,495)(1203,609)
\thicklines \path(1215,495)(1215,599)
\thicklines \path(1227,495)(1227,590)
\thicklines \path(1239,495)(1239,582)
\thicklines \path(1252,495)(1252,574)
\thicklines \path(1264,495)(1264,567)
\thicklines \path(1276,495)(1276,560)
\thicklines \path(1289,495)(1289,554)
\thicklines \path(1301,495)(1301,548)
\thicklines \path(1313,495)(1313,543)
\thicklines \path(1325,495)(1325,538)
\thicklines \path(1338,495)(1338,534)
\thicklines \path(1350,495)(1350,530)
\thicklines \path(1362,495)(1362,526)
\thicklines \path(1375,495)(1375,523)
\thicklines \path(1387,495)(1387,520)
\thicklines \path(1399,495)(1399,518)
\thicklines \path(1411,495)(1411,515)
\thicklines \path(1424,495)(1424,513)
\thicklines \path(1436,495)(1436,511)
\end{picture}
 \end{center}
 \caption[]{\footnotesize 
  Polarized photoabsorption cross section of the electron in
  ${\cal O}(\alpha^2)$ QED.
  Since $\text d\nu/\nu = \text d(\ln\nu)$, the vanishing of (the QED
  contribution to) the integral
  \eqref{wsm-gdh} is reflected by the equality of the shaded areas.
  \label{acm}
 }
\end{figure}
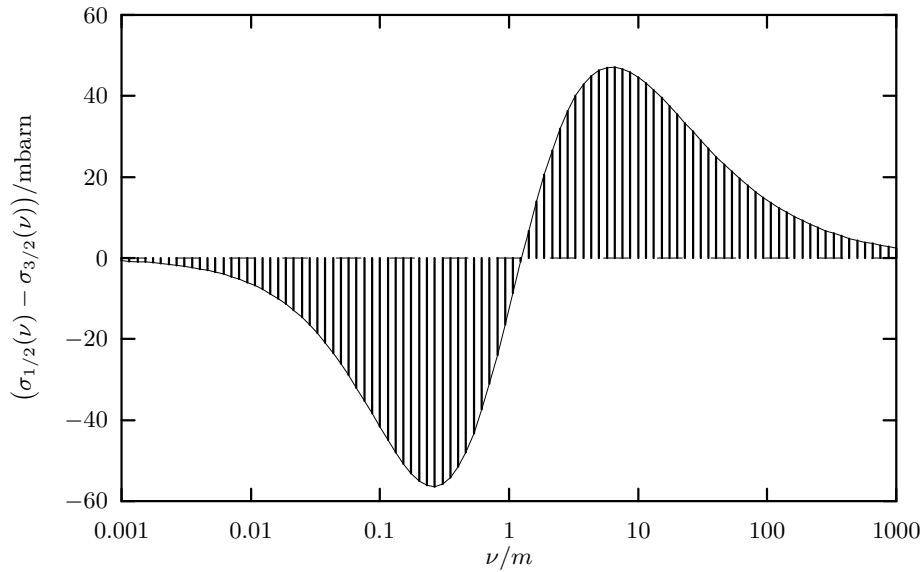

In 1972, Altarelli, Cabibbo, and Maiani \cite{Altarelli72}
investigated the GDH sum rule for the $\mathcal O(\alpha^2)$
Weinberg-Salam model by calculating the forward amplitude $f_2(\nu)$
for the real Compton process and checking explicitely that it obeys an
unsubtracted dispersion relation. The Feynman graphs up to order
$\alpha^2$ are presented in Fig.\ \ref{graphs}.  The tree graphs (a),
as well as the contact graph (e) and the Higgs-exchange graphs (f), do
not contribute to the antisymmetric piece of the Compton amplitude
$T^{\mu\nu}$, thus having no effect on $f_2(\nu)$.  As will be
demonstrated below, in the case of a real photon also the
Z$^0$-exchange graphs of Fig.\ \ref{graphs}(g) vanish. In view of the
fact that the anomalous magnetic moment $\kappa$ of the electron is of
order $\alpha$, the left-hand side of the GDH sum rule \eqref{gdh} is
of order $\alpha^3$, so that to order $\alpha^2$ it reads
\begin{equation} \label{wsm-gdh}
 0 = \int_0^\infty\! \frac{\text d\nu}{\nu}\, \im f_2(\nu).
\end{equation}
This relation is proven in Ref.\ \cite{Altarelli72} (see also Ref.\
\cite{Brodsky95}).  Since $\text d\nu/\nu = \text d(\ln\nu)$, Eq.\
\eqref{wsm-gdh} is reflected by the equality of the shaded areas in
Fig.\ \ref{acm}, which shows the QED contribution to the function
$8\pi\im f_2(\nu) = \sigma_{1/2}(\nu)-\sigma_{3/2}(\nu)$, i.e.\ the
contribution due to the e$^-\gamma$ intermediate states of Fig.\
\ref{graphs}(b).

Here we address the question whether the current-algebra approach to
the GDH sum rule also works within the $\mathcal O(\alpha^2)$
Weinberg-Salam model.  To be precise, we will investigate the validity
of the naive dipole-moment commutator \eqref{naive-comm-me}, and the
legitimacy of the infinite-momentum limit.  We will find that both
assumptions are violated, due to the same Feynman graphs, namely the
Z$^0$-exchange graphs of Fig.\ \ref{graphs}(f). There is a
modification of the finite-momentum GDH sum rule, but this
modification is removed when the infinite-momentum limit is taken,
leading back to the original GDH sum rule \eqref{wsm-gdh}.

We adopt the Bjorken-Johnson-Low (BJL) technique
\cite{Bjorken66,Johnson66} to work out the one-electron matrix element
of the charge-density commutator.  The BJL limit 
\begin{equation} \label{bjl}
 \int\!\text d^3x\, e^{-i\boldsymbol q' \ndot \boldsymbol x}
  \langle p', \tfrac12| [J^0(\boldsymbol x, 0), J^0(0)] |p, \tfrac12\rangle
  = -\lim_{q^{\prime0}\to\infty} q^{\prime0} T^{00}(p,p',q')
\end{equation}
relates the commutator matrix element to the (generally non-forward) 
virtual Compton amplitude
\begin{equation}
 e^2T^{\mu\nu}(p,p',q') = ie^2 \!\int\!\text d^4x\, e^{iq'\ndot x}
  \langle p', \tfrac12| \text T J^\mu(x) J^\nu(0) | p, \tfrac12 \rangle,
\end{equation}
which we study perturbatively.
All polynomials in $q^{\prime0}$, the so-called seagulls, have to be dropped in
this procedure.
A typical seagull is presented by the Feynman graph of Fig.\ \ref{graphs}(e),
which is completely independent of the photon momenta $q$ and $q'$.

We found that to order $\alpha^2$, only the Z$^0$-exchange graphs of
Fig.\ \ref{graphs}(f) contribute.  This may not be surprising, since
these fermion triangle graphs are responsible for the famous
Adler-Bell-Jackiw anomaly \cite{Adler69a,Bell69}.  Details of our
calculation are given in the appendix.  The result for
the matrix element of the dipole-moment commutator is non-vanishing,
\begin{equation} \label{wsm-comm-me}
 \langle p', \tfrac12 | [D^+(0), D^-(0)] | p, \tfrac12 \rangle =
  (2\pi)^3\, 2p^0\, \delta^3(\boldsymbol p' - \boldsymbol p)\,
  \frac\alpha\pi\, \frac{G_{\text F}}{\sqrt2},
\end{equation}
in contrast to the naive assumption \eqref{naive-comm-me}.  Re-inspecting
now the derivation of the finite-momentum GDH sum rule presented
in the previous section, we infer a modification given by
\begin{equation} \label{wsm-pre-gdh}
 0 = \alpha\,\frac{G_{\text F}}{\sqrt2} -
  \lim_{p^0\to\infty} \int_0^\infty\! \frac{\text d\nu}\nu\,
  8\pi \im f_2\!\left(\nu,\frac{m^2\nu^2}{(p^0)^2}\right).
\end{equation}
Obviously, the GDH sum rule \eqref{wsm-gdh} would be violated if limit
and integration were interchangeable. In the following we will show that due
to the same Feynman graphs that gave rise to the anomalous commutator
\eqref{wsm-comm-me}, dragging the limit inside the integral results in
a second modification that cancels the first one. This means that the naive
infinite-momentum limit is illegitimate here!

\begin{figure}[tb]
 \begin{center}
  \input{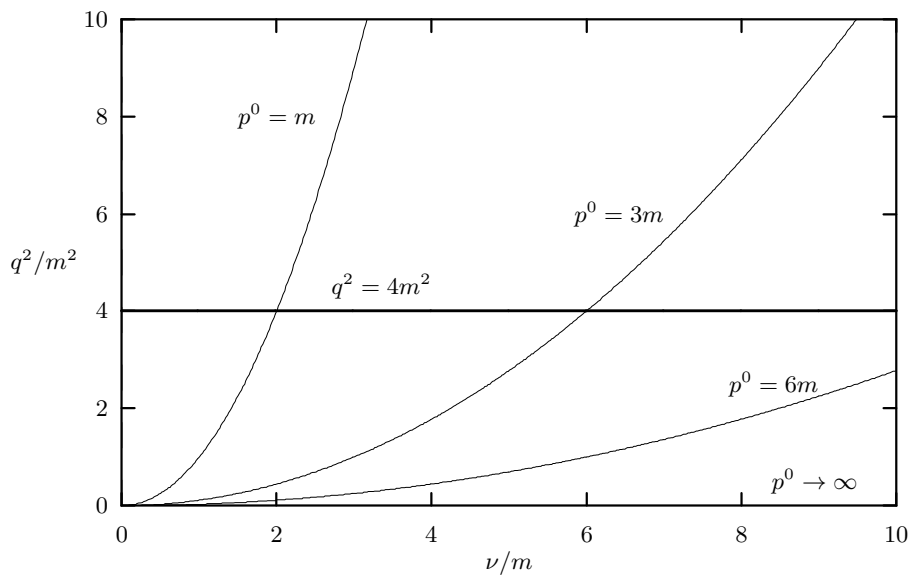}
 \end{center}
 \caption{\footnotesize
  The integration path of the finite-momentum GDH sum rule of the 
  $\mathcal O(\alpha^2)$ Weinberg-Salam model for electron energy $p^0=m$,
  $p^0=3m$, and $p^0=6m$.  For any finite value of $p^0$ the integration
  passes the two-electron threshold $q^2=4m^2$, picking up the constant on
  the right-hand side of Eq.\ \eqref{limint}.
  \label{q2nu-wsm}}
\end{figure}

To prove this assertion, we calculated the Z$^0$-exchange contribution
to the amplitude $f_2(\nu,q^2)$.  Some details are given in the
appendix.  The result is independent of the photon energy $\nu$,
\begin{equation} \label{f2z}
 f_2^{(\text Z)}(\nu,q^2) = \frac\alpha{4\pi^2}\,
  \frac{G_{\text F}}{\sqrt2}\, f(q^2),
\end{equation}
where the function $f(q^2)$ can be given explicitely. As expected, it
exhibits a branch-point singularity at the two-electron threshold
$q^2=4m^2$.  Below this threshold one has
\begin{align}  \label{re}
 \re f(q^2) &= \frac{4m^2}{\sqrt{(4m^2-q^2)q^2}}
  \arccot{\sqrt{\frac{4m^2}{q^2}-1}} \,-\, 1,\\ 
  \im f(q^2) &= 0,
\end{align}
while for $q^2 > 4m^2$,
\begin{align}
 \re f(q^2) &= \frac{4m^2}{\sqrt{(q^2-4m^2)q^2}}
  \artanh{\sqrt{1-\frac{4m^2}{q^2}}} \,-\, 1,\\
  \im f(q^2) &= \frac{2\pi m^2}{\sqrt{(q^2-4m^2)q^2}}. \label{im}
\end{align}%
\begin{figure}[tb]
 \begin{center}
  \input{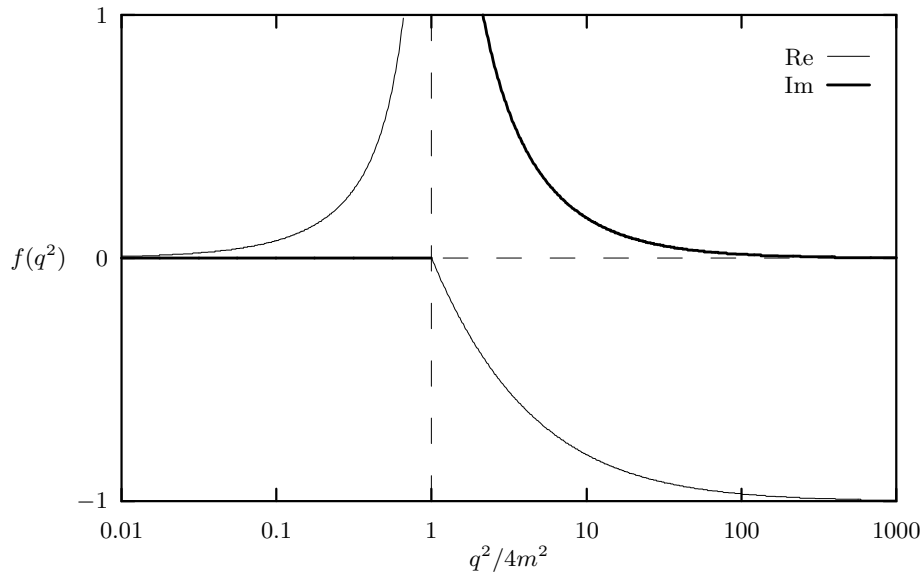}
 \end{center}
 \caption{\footnotesize
  Real part (light line) and imaginary part (heavy line)
  of the function $f(q^2)$ occuring in the Z$^0$-exchange contribution
  $f_2^{(\text Z)}(\nu,q^2) \propto f(q^2)$
  to the virtual forward Compton amplitude of the
  $\mathcal O(\alpha^2)$ Weinberg-Salam model.
  \label{fq2}}
\end{figure}%
This function is depicted in Fig.\ \ref{fq2}. For real Compton scattering,
$q^2 = 0$, there is no contribution,
\begin{equation}
 f_2^{(\text Z)}(\nu,0) = 0,
\end{equation}
and hence
\begin{equation} \label{intlim}
 \int_0^\infty\! \frac{\text d\nu}\nu \im f_2^{(\text Z)}(\nu,0) = 0.
\end{equation}
However, for timelike photon virtualities above the two-electron threshold,
one has a non-vanishing imaginary part.  We infer from eqs.\
\eqref{f2z} and \eqref{im},
\begin{equation} \label{limint}
 \int_0^\infty\! \frac{\text d\nu}\nu\,
  8\pi\im f_2^{(\text Z)}\!\left(\nu,\frac{m^2\nu^2}{(p^0)^2}\right) =
  \alpha\, \frac{G_{\text F}}{\sqrt2},
\end{equation}
which is \emph{independent} of the electron energy $p^0$.
As can be seen in Fig.\ \ref{q2nu-wsm}, the contour of the integration
in Eq.\ \eqref{limint} passes the singularity line at $q^2=4m^2$ for
any finite value of $p^0$.%

The crucial observation now is that interchanging the $\nu$ integration
and the $p^0\to\infty$ limit leads to an additive constant
\begin{equation}
 \left(\lim_{p^0\to\infty}\int_0^\infty\!\frac{\text d\nu}\nu -
  \int_0^\infty\!\frac{\text d\nu}\nu\lim_{p^0\to\infty}\right)
  8\pi\im f_2\!\left(\nu,\frac{m^2\nu^2}{(p^0)^2}\right) =
  \alpha\,\frac{G_{\text F}}{\sqrt2},
\end{equation}
due to the Z$^0$-exchange contributions \eqref{intlim} and \eqref
{limint}. Combined with the finite-momentum GDH sum rule
\eqref{wsm-pre-gdh}, this leads back to the undisturbed GDH sum rule
\eqref {wsm-gdh}.

We remark that if quarks are included into the model, then
the customary effect of anomaly cancellation removes the modification
of the charge-density commutator as well as the one due to the
infinite-momentum limit.

\section{Summary and conclusion}

We presented a derivation of the GDH sum rule from the equal-times
commutator of electric charge densities.  Our derivation exhibits the
infinite-momentum limit as its last step.  The \emph{finite-momentum
GDH sum rule} \eqref{fmgdh} gives the form that the sum rule
takes without performing the infinite-momentum limit.  We emphasized
the fact that in principle, taking the infinite-momentum limit
could give rise to a modification of the GDH sum rule, and that
current algebra alone cannot tell whether such a modification is
present or not.

To get a feeling for what can happen when taking the infinite-momentum
limit, we considered virtual Compton scattering off the electron
within the Weinberg-Salam model of weak interactions to order
$\alpha^2$.  We found that in this model the infinite-momentum
limit does indeed give rise to a certain finite modification of the
GDH sum rule, which comes together with (and is cancelled by)
another modification due to an anomalous charge-density commutator.

The coincidence of these modifications leads us to the conclusion that
any proposal of a modification of the GDH sum rule suggested from an
anomalous charge-density commutator that has no regard to the legitimacy
of the infinite-momentum limit (such as Ref.\ \cite{Chang94a}) 
is to be seriously doubted!

\begin{appendix}

\section{Z$^{\boldsymbol 0}$ exchange}

This appendix shall describe the calculation of the dipole-moment
commutator matrix element \eqref{wsm-comm-me} from the BJL formula
\eqref{bjl}, originating, to order $\alpha^2$, from the Z$^0$-exchange
graphs of Fig.\ \ref{graphs}(f).  Also, the calculation of the
contribution \eqref{f2z} of these graphs to the forward amplitude
$f_2(\nu,q^2)$ is illustrated.
The relevant Feynman integrals have been worked out by Rosenberg
\cite {Rosenberg63} and were discussed further by Adler \cite {Adler69a}.

We are concerned with the (generally non-forward) Compton amplitude
\begin{equation}
 e^2T^{\mu\nu}(p,p',q',s,s') = ie^2 \!\int\!\text d^4x\, e^{iq'\ndot x}
  \langle p', s'| \text T J^\mu(x) J^\nu(0) | p, s \rangle.
\end{equation}
The contribution from the Z$^0$-exchange graphs reads
\begin{equation} \label{tz}
 T^{\mu\nu}_{(\text Z)} = - \frac{M_{\text Z}^2G_{\text F}}{2\sqrt2}
  R^{\mu\nu\rho}(q,q')
  \frac{-g_{\rho\sigma}+(p'-p)_\rho(p'-p)_\sigma/M_{\text Z}^2}
   {(p'-p)^2-M_{\text Z}^2}\,
  \bar u(p',s') \gamma^\sigma\gamma_5 u(p,s),
\end{equation}
with
\begin{equation} \label{r-def}
 R^{\mu\nu\rho} = \int\!\frac{\text d^4k}{(2\pi)^4}
  \tr\left(\gamma^\mu\frac i{\fslash k-m+i\epsilon}
  \gamma^\nu\frac i{\fslash k-\fslash q-m+i\epsilon}  
  \gamma^\rho\gamma_5\frac i{\fslash k-\fslash q'-m+i\epsilon} \right)
  + \genfrac{\{}{\}}{0pt}{}{\mu\leftrightarrow\nu}{q\leftrightarrow-q'}.
\end{equation}
The constant $M_{\text Z}^2G_{\text F}/2\sqrt2$ is due to the coupling of the 
Z$^0$ to the electron lines, $m$ and $M_{\text Z}$ are the masses of e$^-$
and Z$^0$, respectively. The momentum four-vector $q$ of the incoming
photon is fixed by $p+q=p'+q'$.  This has to be kept in mind when the 
$q^{\prime0}\to\infty$ limit is taken.

The triangle loop integral \eqref{r-def} can be cast into the form
\begin{align} \label{r-res}
 R^{\mu\nu\rho} & = \epsilon^{\mu\nu\rho\alpha} q_\alpha R_1
  + \epsilon^{\nu\rho\alpha\beta}q'_\alpha q_\beta
  (q^\mu R_2 + q^{\prime\mu} R_3)
  \notag\\
 &\quad + \epsilon^{\mu\nu\rho\alpha} q'_\alpha R'_1
  + \epsilon^{\mu\rho\alpha\beta}q'_\alpha q_\beta
  (q^{\prime\nu} R'_2 + q^\nu R'_3),
\end{align}
where $R_{1,2,3}$ are functions of the Lorentz invariants $q^2$,
$q^{\prime2}$, and $q\ndot{q'}$.  From crossing symmetry, the primed
quantities are given by $R'_{1,2,3}(q^2,q^{\prime2},q\ndot{q'}) \equiv
R_{1,2,3}(q^{\prime2},q^2,q\ndot{q'})$.  Gauge invariance
$q_{\nu}T_{(\text Z)}^{\mu\nu}=0$, $q'_{\mu}T_{(\text Z)}^{\mu\nu}=0$,
imposes the condition
\begin{equation}  \label{gaugeinv}
 R_1 + q\ndot{q'}R_2 + q^{\prime2}R_3 = R'_1 + q\ndot{q'}R'_2 + q^2R'_3 = 0.
\end{equation}
The crucial point of Ref.\ \cite {Rosenberg63} is the observation that
the functions $R_{2,3}$ are finite, while the formally divergent function
$R_1$ can be fixed by the gauge-invariance condition \eqref{gaugeinv}.
The result is given by the Feynman-parameter integrals
\begin{equation} \label{r123}
 R_{1,2,3}^{(\prime)}(q,q') =
  \frac{i}{\pi^2}\!\int_0^1\!\text dx \int_0^{1-x}\!\text dy\,
  \frac{N^{(\prime)}_{1,2,3}}
   {x(1-x)\,q^2 + y(1-y)\,q^{\prime2} - 2xy\,q\ndot q' - m^2 + i\epsilon},
\end{equation}
with the numerators
\begin{equation} \label{n123}
 \begin{split}
  &N_1 = x(1-x)\,q^{\prime2}-xy\,q\ndot q', \quad N_2 =  xy,
   \quad N_3 = -x(1-x), \\
  &N'_1 = y(1-y)\,q^2-xy\,q\ndot q', \quad N'_2 =  xy,
   \quad N'_3 = -y(1-y).
 \end{split}
\end{equation}

\subsection{Anomalous commutator} \label{anom-comm}

To compute the matrix element of the dipole-moment commutator, we need
the $q^{\prime0}\to\infty$ limit of the time-time component $T_{(\text
Z)}^{00}$.  We have, from Eq.\ \eqref{r-res},
\begin{equation}
 \lim_{q^{\prime0}\to\infty} q^{\prime0} R^{00\rho} =
  \epsilon^{0\rho\alpha\beta} q'_\alpha q_\beta
  \lim_{q^{\prime0}\to\infty}
  \left((q^{\prime0})^2(R_2+R'_2+R_3+R'_3)\right).
\end{equation}
Carrying out the $q^{\prime0}\to\infty$ limit in the explicite
formulae \eqref{r123} and \eqref{n123}, this reduces to
\begin{equation}  \label{r2r3}
 \lim_{q^{\prime0}\to\infty} q^{\prime0} R^{00\rho} =
  -\frac i{2\pi^2}\, \epsilon^{0\rho\alpha\beta} q'_\alpha q_\beta.
\end{equation}
Inserting this into Eq.\ \eqref{tz} and going back to the BJL formula
\eqref{bjl}, it is a trivial matter to work out the matrix element of the
charge-density commutator,
\begin{multline}
 \langle p', \tfrac12| [J^0(\boldsymbol x, 0), J^0(0)] |p, \tfrac12\rangle = \\
  -\epsilon^{ijk}\,\frac{M_{\text Z}^2G_{\text F}}{4\pi^2\sqrt2}\,
  \frac{(p' - p)^i\,
  \bar u(p',\tfrac12)\gamma^j\gamma_5u(p,\tfrac12)\,
  \nabla^k\delta^3(\boldsymbol x)}{(p'-p)^2 - M_{\text Z}^2}.
\end{multline}
The matrix element \eqref{wsm-comm-me} of the dipole-moment
commutator is now obtained by using translational invariance and the properties
of the helicity spinor $u(p,\frac12)$.

\subsection{Infinite-momentum limit} \label{iml}

On the other hand, we want to compute the Z$^0$-exchange contribution
to the forward Compton amplitude $f_2(\nu,q^2)$. To this end, we have
to specialize eqs.\ \eqref{tz} and \eqref{r-res} to the case $p=p'$,
$q=q'$, $s=s'$, giving
\begin{equation}  \label{r1}
 T_{(\text Z)}^{\mu\nu}(p,q,s) = m\,\frac{G_{\text F}}{\sqrt2}\,
 \epsilon^{\mu\nu\rho\sigma} q_\rho s_\sigma (R_1+R'_1).
\end{equation}
Recalling the invariant decomposition \eqref{inv-dec}, we notice that Z$^0$
exchange contributes to $A_1(\nu,q^2)$ only,
\begin{align}
 A_1^{(\text Z)}(\nu,q^2)
  &= -\frac{m^2}{\pi^2}\,\frac{G_{\text F}}{\sqrt2}
  \int_0^1\!\text dx \!\int_0^{1-x}\!\text dy\,
  \frac{(x+y)(1-x-y)\,q^2}{(x+y)(1-x-y)\,q^2-m^2+i\epsilon} \notag\\
 &= \frac{m^2}{2\pi^2}\,\frac{G_{\text F}}{\sqrt2}\, f(q^2),
\end{align}
where
\begin{align}  \label{zint}
 f(q^2) =
  -2 \int_0^1\!\text dz\, \frac{z^2(1-z)\,q^2}{z(1-z)\,q^2-m^2+i\epsilon}.
\end{align}
This implies Eq.\ \eqref{f2z} via relation \eqref{f2nuq2}.
The $z$ integration can be performed explicitely, giving
formulae \eqref{re}--\eqref{im}.

Finally, it is perhaps worth noting that the exact cancellation of the
two modifications of the GDH sum rule can be traced back to the
gauge-invariance condition \eqref {gaugeinv}, which relates the function
$R_1$ occuring in Eq.\ \eqref{r1} to the functions
$R_{2,3}$ of Eq.\ \eqref{r2r3}.

\end{appendix}

\bibliographystyle{prsty}
\bibliography{particle}

\end{document}